\documentclass[conference]{IEEEtran}
\IEEEoverridecommandlockouts
\usepackage{cite}
\usepackage{amsmath,amssymb,amsfonts}
\usepackage{algorithmic}
\usepackage{graphicx}
\usepackage{textcomp}
\usepackage{xcolor}
\usepackage[a4paper, total={184mm,239mm}]{geometry}

\usepackage[linesnumbered,ruled]{algorithm2e}
\usepackage{setspace}
\usepackage{multirow}
\usepackage{subcaption}
\usepackage{array}
\usepackage{graphicx}
\usepackage{threeparttable}

\def\BibTeX{{\rm B\kern-.05em{\sc i\kern-.025em b}\kern-.08em
		T\kern-.1667em\lower.7ex\hbox{E}\kern-.125emX}}
\begin{document}
	
	\title{Mr.TPL: A Method for Multi-Pin Net Router in Triple Patterning Lithography
		\thanks{Corresponding author: Hailong Yao and Weiqing Ji.}
	}
	
	\author{\IEEEauthorblockN{Chengkai Wang\textsuperscript{1}, Weiqing Ji\textsuperscript{1}, Mingyang Kou\textsuperscript{1}, Zhiyang Chen\textsuperscript{2}, Fei Li\textsuperscript{3} and Hailong Yao\textsuperscript{1}}
		\IEEEauthorblockA{\textsuperscript{1}\textit{University of Science and Technology Beijing}, Beijing, China \\
			\textsuperscript{1}\textit{Tsinghua University}, Beijing, China\\
			\textsuperscript{3}\textit{Empyrean Technology Co., Ltd.}, Beijing, China\\
			chengkaiwang@xs.ustb.edu.cn, weiqingji@ustb.edu.cn, kmy@ustb.edu.cn,\\ chenzhiy21@mails.tsinghua.edu.cn, lifei2@empyrean.com.cn, hailongyao@ustb.edu.cn}
		}
	
	\maketitle
	
	\begin{abstract}
		Triple patterning lithography (TPL) has been recognized as one of the most promising solutions to print critical features in advanced technology nodes. A critical challenge within TPL is the effective assignment of the layout to masks. Recently, various layout decomposition methods and TPL-aware routing methods have been proposed to consider TPL. However, these methods typically result in numerous conflicts and stitches, and are mainly designed for 2-pin nets. This paper proposes a multi-pin net routing method in triple patterning lithography, called Mr.TPL. Experimental results demonstrate that Mr.TPL reduces color conflicts by 81.17\%, decreases stitches by 76.89\%, and achieves up to 5.4$\times$ speed improvement compared to the state-of-the-art TPL-aware routing method.
	\end{abstract}
	
	\begin{IEEEkeywords}
		triple patterning aware routing, detail routing, multiple patterning lithography
	\end{IEEEkeywords}

	\section{Introduction}
	
	Triple patterning lithography has been recognized as one of the most promising solutions to print critical features in advanced technology nodes\cite{areWeReady}. The primary challenge of TPL lies in the efficient decomposition of design layouts, ensuring that patterns are correctly assigned to different masks.
	
	In recent years, various layout decomposition methods\cite{openmpl, LRSDP, LDBS} and TPL-aware routing methods\cite{dac2012, triad, nostitch} have been proposed to consider TPL. Li et al. developed a layout decomposition method, called OpenMPL\cite{openmpl}, which does not change the layout but assigns features that are close to each other to different masks. Zhang et al. proposed a dedicated low-rank SDP algorithm for MPL decomposition with augmented Lagrangian relaxation and Riemannian optimization \cite{LRSDP}. However, since the layout patterns remain unchanged, existing layout decomposition methods inevitably lead to unsolvable color conflict issues. As shown in Fig~\ref{figure1}(a), the four patterns are too close to each other, making it impossible to assign three colors in a rule-compliant manner, thus introducing an unresolvable color conflict. Fig~\ref{figure1}(b) is a reasonable solution.
	
	Ma et al. introduced a TPL aware routing method that modifies the routing grid, splits each vertice into 12 vertices to represent different masks and directions, and then applies routing\cite{dac2012}. Lin et al. proposed a TPL aware routing method, which uses a token graph-embedded conflict graph\cite{triad}. On the one hand, most TPL-aware routing methods have a very high time complexity, typically running 3-10 times slower. On the other hand, these methods are primarily designed for 2-pin connections, resulting in an excessive number of stitches when applied to multi-pin nets, as shown in Fig~\ref{figure1}(c). This occurs because 2-pin methods cannot dynamically adjust the already-colored paths when connecting multiple pins. Fig~\ref{figure1}(d) is a reasonable solution.
	
	To address the above issue, we propose a novel detailed routing method, called Mr.TPL, that considers TPL during the detailed routing phase, specifically targeting multi-pin nets. Our major contributions are as follows:
	
	\begin{figure}[t]
		\centering
		\begin{subfigure}{0.49\linewidth}
			\centering
			\includegraphics[width=1\linewidth]{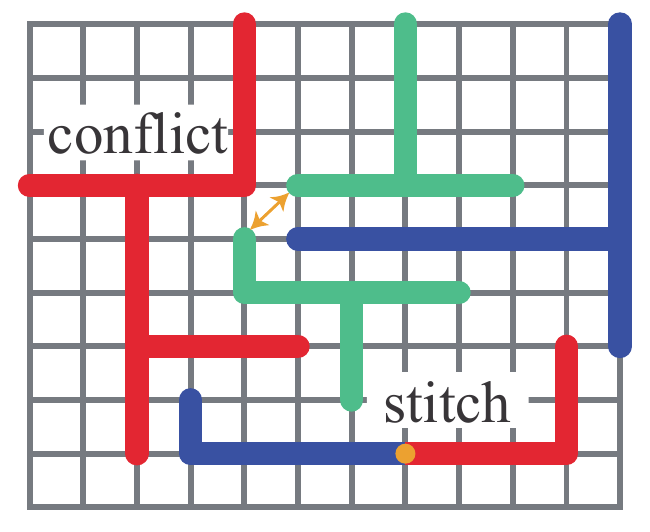}
			\caption{}
		\end{subfigure}
		\begin{subfigure}{0.49\linewidth}
			\centering
			\includegraphics[width=1\linewidth]{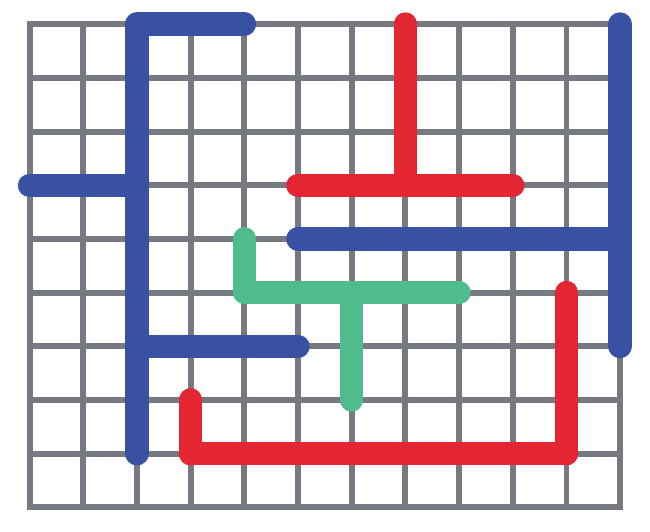}
			\caption{}
		\end{subfigure}
		\begin{subfigure}{0.49\linewidth}
			\centering
			\includegraphics[width=1\linewidth]{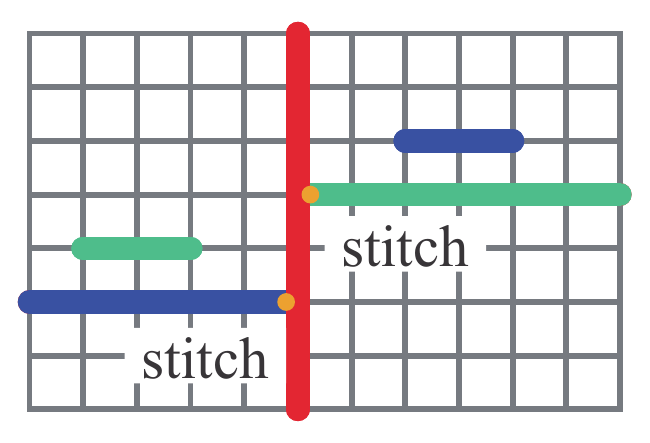}
			\caption{}
		\end{subfigure}
		\begin{subfigure}{0.49\linewidth}
			\centering
			\includegraphics[width=1\linewidth]{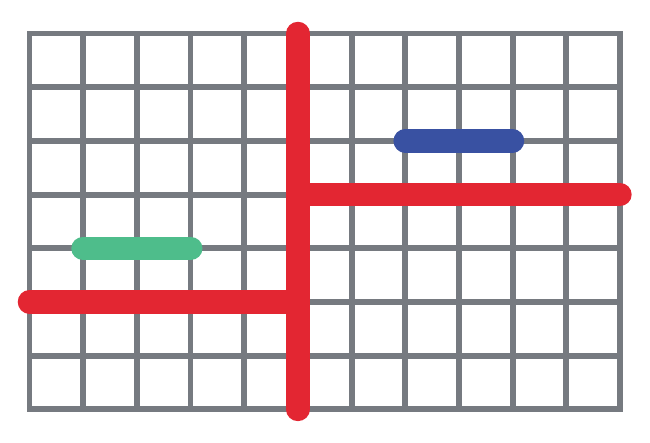}
			\caption{}
		\end{subfigure}
		\caption{Examples of TPL solution. (a) The routing solution with an unsolvable color conflict and a stitch. (b)(d) The routing solution without color conflicts or stitches. (c) Stitches caused by 2-pin TPL routing method.}
		\label{figure1}
	\end{figure}
	
	\begin{itemize}
		\item We propose a search algorithm based on detail routing that better handles the TPL issues in multi-pin nets. To the best of our knowledge, this is the first work on the routing problem of multi-pin nets in TPL.
		\item We propose a detailed routing method that considering TPL in the routing stage of multi-pin nets, which can greatly reduce the number of color conflicts and stitches.
		\item We develop a set-based color state merging and vertice coloring method that allows routing path to be in triple color candidate states at the same time during the search process, rather than a determined single color mask.
		\item Compared with the state-of-the-art TPL-aware 
		routing method \cite{dac2012}, Mr.TPL reduces color conflicts by 81.17\% and stitches by 76.89\%, while achieving a speedup of up to 5.4$\times$ on ISPD benchmarks.
	\end{itemize}
	
	The rest of this paper is organized as follows. Section 2 provide the preliminary and problem formulation of the TPL-aware detail routing. Section 3 provides a overall workflow of the proposed method. Section 4 elaborates on the various stages of the method. Section 5 presents experimental results, followed by Section 6, which offers conclusions based on the findings.

	\section{Preliminary and Problem Formulation}
	
	In this section, we first introduce the TPL aware routing. Then, we formulate the TPL aware routing problem for multi-pin nets.
	
	\subsection{TPL-aware Routing}
	
	Triple Patterning Lithography (TPL) is a photolithography technique used in semiconductor manufacturing to overcome the limitations of traditional lithography at advanced technology nodes \cite{areWeReady}. By splitting a design layer into three separate masks, TPL enables the printing of smaller features, improving resolution. However, TPL introduces significant challenges in mask assignment and routing, as conflicts between features must be avoided to ensure manufacturability. TPL-aware routing integrates mask assignment considerations into the routing stage, addressing mask assignment earlier in the design process. 
	
	Fig~\ref{figure1} shows examples of routing solution considering TPL. As shown in Fig~\ref{figure1}(a), when the distance between patterns on a layout falls below a predefined threshold, these patterns must be assigned to separate masks to avoid conflicts. If such conflicts arise, one way to resolve them is by assigning the layout to triple masks, introducing a stitch at the boundary where the color changes. Conflicts can lead to printing failures in the metal layout, while stitches may reduce the overall yield. The primary goal of TPL is to minimize both the number of conflicts and stitches. Fig~\ref{figure1}(b) shows the routing solution without conflicts or stitches.
	
	\subsection{Problem Formulation}
	
	
	The TPL aware routing problem for multi-pin nets can be formulated as follows.
	
	
	\textbf{Given:} (1) Layout, including the distribution of pre-placed standard cells, macros, obstacles, and ports.
	(2) The netlist, which describes the connections between components in the layout.
	(3) Design rules, including the design rules mentioned in ISPD contests\cite{ispd18} and different mask spacing $D_{color}$.
	
	\textbf{Goal:} A routed and colored layout that minimizes the weighted sum of conflict cost and stitch cost.
	
	%
	%
	%
	
	\section{Overview}
	
	\begin{figure}[htbp]
		\centering
		\includegraphics[width=1\linewidth]{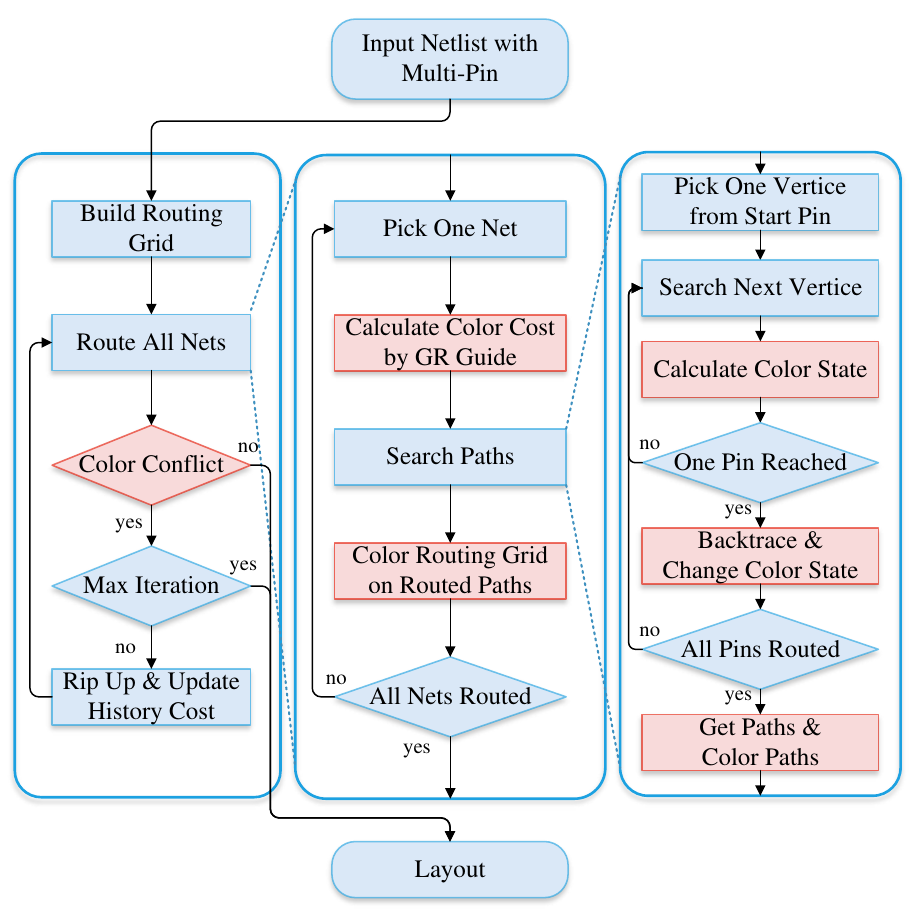}
		\caption{Overall flow of Mr.TPL}
		\label{figure2}
	\end{figure}
	
%
%
	
	Fig~\ref{figure2} illustrates the overall flow of Mr.TPL. Before routing begins, the necessary data must be fully loaded including layout information, netlist, and global routing results. We also construct the routing grid at this stage. Once the data is ready, the routing process commences. After routing is completed, conflict detection is performed. If conflicts are identified and the maximum iteration limit has not been reached, the process proceeds with rip-up, reroute and updates the history cost accordingly.
	
	The central column details the routing process. At the start of routing, a net is selected for routing. The color cost is calculated with routing regions based on global routing (GR) guide. The path-searching phase follows, during which a colored path is identified. This path is then used to mark the routed vertices and their associated colors on the routing grid. The process iterates until all nets are successfully routed.
	
	The third column describes the detailed steps of the search path process. A pin is randomly selected from the net as the starting point for path searching. Also a vertice is selected within the coverage of the start pin. During searching to the next vertice, the color state of the vertice is calculated. This step is repeated until the next pin is reached. Subsequently, a backtrace phase and color state change is performed. The process repeats until all pins are routed. Finally, the paths are generated and colored.
	
	\begin{figure*}[htbp]
		\centering
		\begin{subfigure}{0.32\linewidth}
			\centering
			\includegraphics[width=1\linewidth]{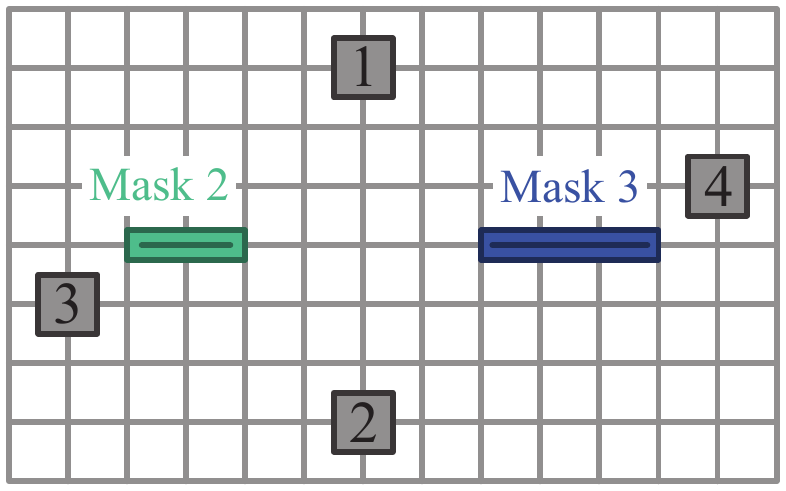}
			\caption{Before routing the 4-pin net.}
		\end{subfigure}
		\begin{subfigure}{0.32\linewidth}
			\centering
			\includegraphics[width=1\linewidth]{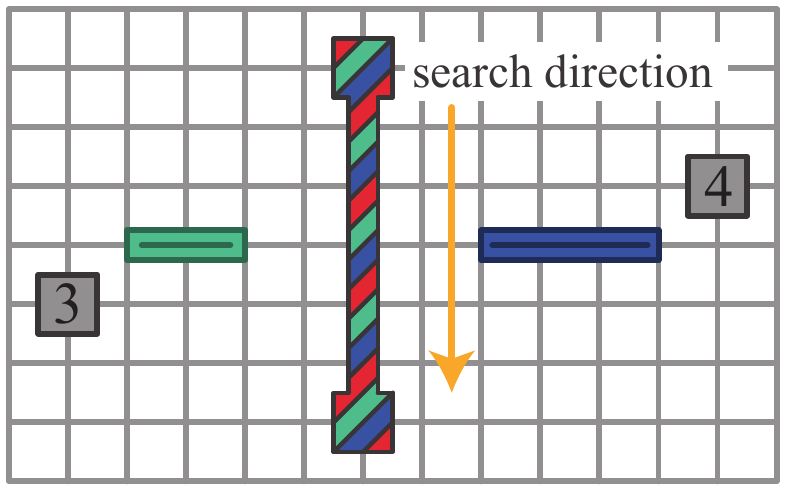}
			\caption{Routed path with color state 111.}
		\end{subfigure}
		\begin{subfigure}{0.32\linewidth}
			\centering
			\includegraphics[width=1\linewidth]{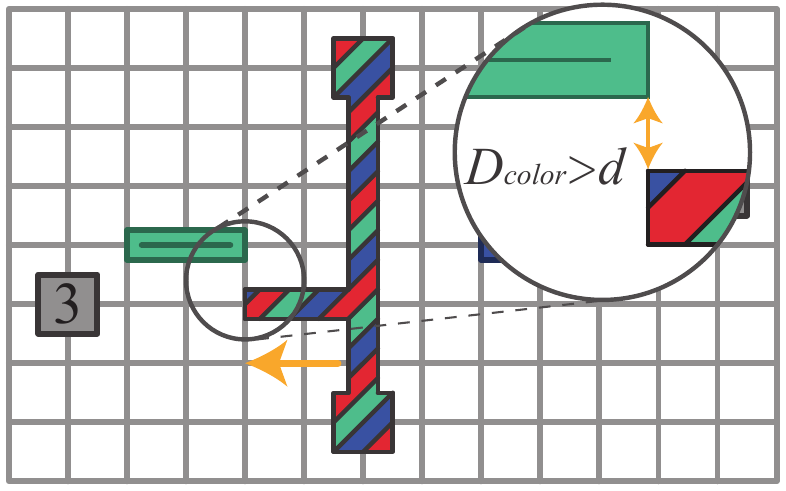}
			\caption{Conflict detection during searching.}
		\end{subfigure}
		\begin{subfigure}{0.32\linewidth}
			\centering
			\includegraphics[width=1\linewidth]{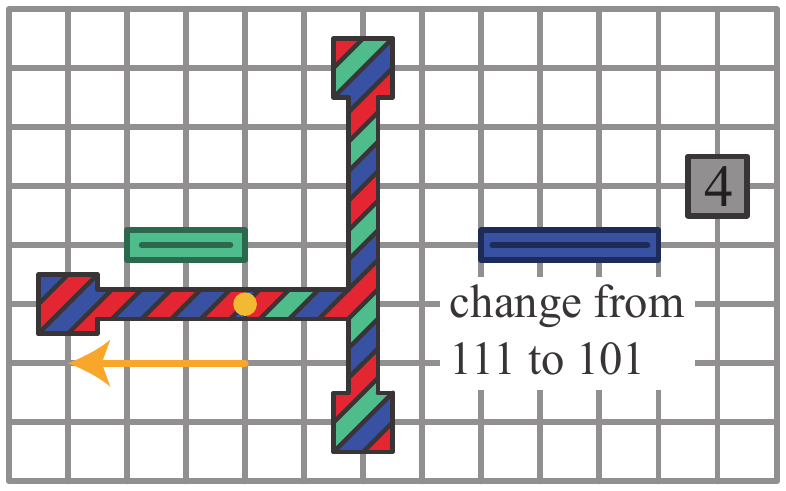}
			\caption{Change color state during searching.}
		\end{subfigure}
		\begin{subfigure}{0.32\linewidth}
			\centering
			\includegraphics[width=1\linewidth]{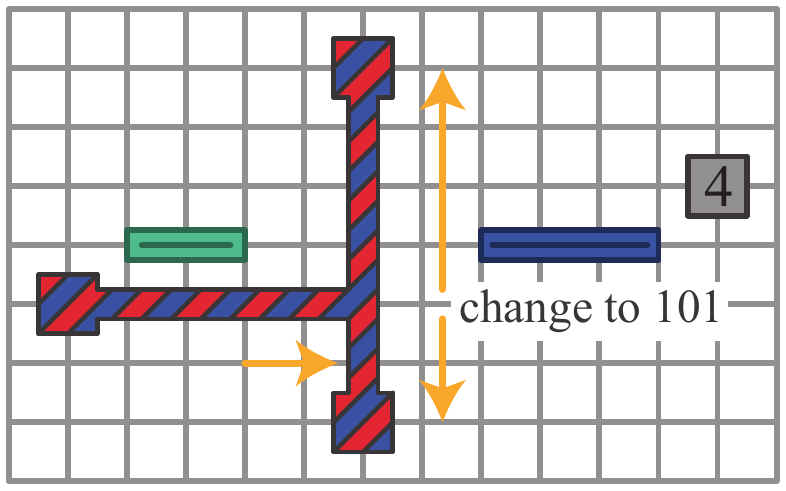}
			\caption{Backtrace and change color state.}
		\end{subfigure}
		\begin{subfigure}{0.32\linewidth}
			\centering
			\includegraphics[width=1\linewidth]{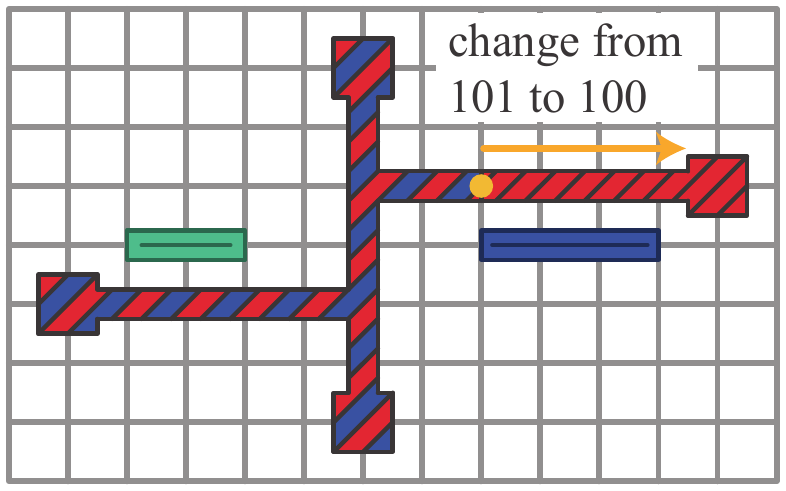}
			\caption{Change color state during searching.}
		\end{subfigure}
		\begin{subfigure}{0.32\linewidth}
			\centering
			\includegraphics[width=1\linewidth]{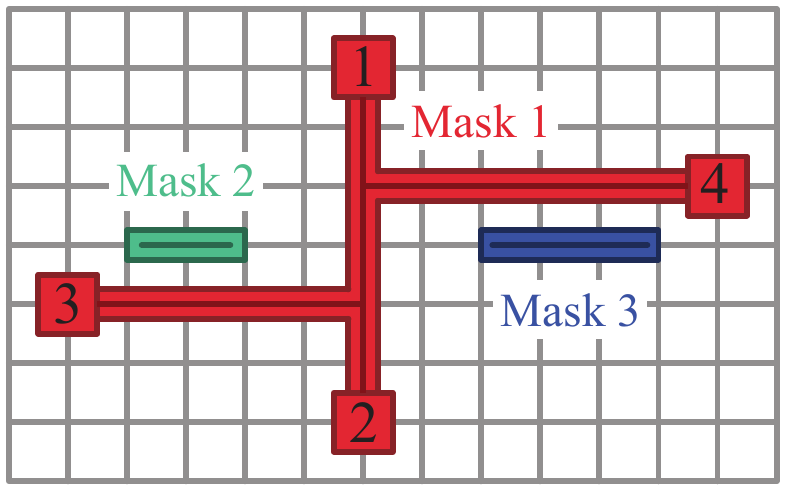}
			\caption{Backtrace and final layout.}
		\end{subfigure}
		\begin{subfigure}{0.43\linewidth}
			\centering
			\includegraphics[width=1\linewidth]{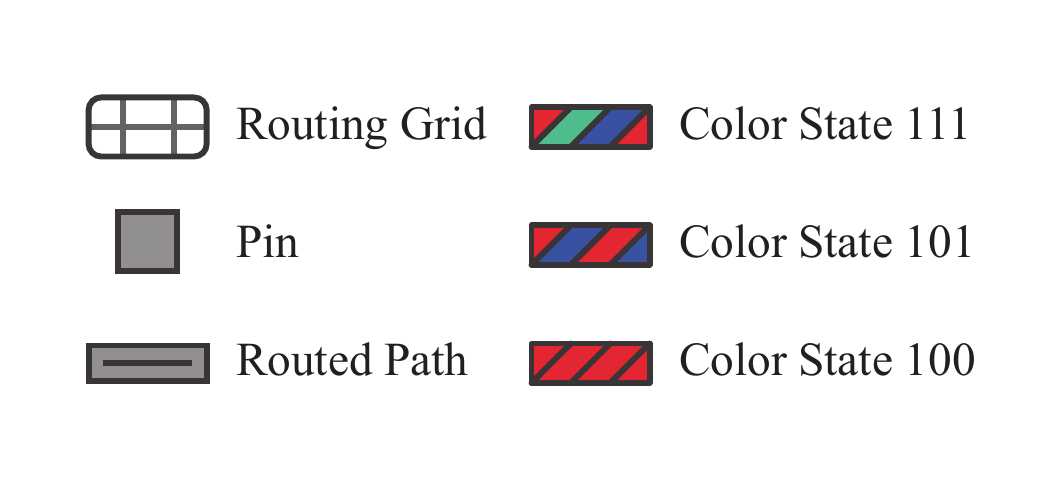}
			\subcaption*{}
		\end{subfigure}
		\caption{Example of Mr.TPL.}
		\label{figure3}
	\end{figure*}
	
	\section{Method}
	
	In this section, we will introduce details of Mr.TPL. First, we will introduce the color state and how the routing grid is constructed. Then, we will explain how the cost is calculated when color is introduced. Next, we will describe how the color state of a vertice is determined during the search process and how we handle the adding of stitches when necessary. Finally, we will detail our approach to assigning colors to vertices during the backtracing stage. Fig~\ref{figure3} shows an example of Mr.TPL.
	
	\subsection{Color State}
	To tackle the challenge of color representation in the detailed routing process, we introduce a property called \textit{color state} for each wire segment, which represents a part of the wire. Now we have the definition here.
	
	
	\textit{\textbf{Definition 1:}} In Mr.TPL, a \textit{\textbf{color state}} refers to the preparatory assignment of different colors to the routing segments on the same metal layer, ensuring compliance with design rules while minimizing color conflicts and associated manufacturing costs.
	
	\hspace*{\fill}
	
	\begin{table}[!t]
		\caption{color state description}
		\begin{center}
			\begin{tabular}{|c|l|}
				\hline
				Encode&Description\\
				\hline
				000&none color is allowed\\
				100&only red is allowed\\
				010&only green is allowed\\
				001&only blue is allowed\\
				110&red and green are allowed\\
				101&red and blue are allowed\\
				011&green and blue are allowed\\
				111&all colors are allowed\\
				\hline
			\end{tabular}
		\end{center}
	\end{table}

	This property plays a critical role in managing the triple patterning process and ensuring the correct application of masks. The color state is represented as a 3-bit unsigned integer, effectively encapsulating the necessary information. Each bit corresponds to one of the three masks—red, green, and blue. For example, a binary value of 100 represents red, 010 represents green, and 001 represents blue.
	
	During the routing search process, a wire segment's color state can simultaneously include multiple mask options. When the color state includes three colors, it implies that assigning any of these colors will incur the same cost. For instance, a color state of 110 indicates that the wire segment can be assigned to either the red or green mask, but not the blue.
	
	At the end of the backtracing phase, a single mask is selected for each wire segment. During this phase, the color state of each segment gradually converges to a single color, ensuring an optimal routing solution. This approach allows for flexible and efficient color management throughout the routing process, significantly enhancing the results in triple patterning lithography.

	\subsection{Graph Construction and Cost Calculation}
	
	In the initialization phase of triple patterning detailed routing, it is essential to construct the graphical representation of the routing grid. We model it as an undirected graph \( G = (V, E) \), where \( V \) denotes the set of vertices representing the intersections or corners within the vertice, and \( E \) represents the set of edges indicating the connectivity between vertices. To incorporate triple patterning manufacturing considerations, we augment each edge \( e \in E \) with an additional attribute, referred to as the \textit{color state}, denoted by \( c(e) \). The color state \( c(e) \) takes values from a predefined finite set \( C = \{c_{100}, c_{010}, \dots, c_{111}\} \), which corresponds to the number of color states used in the process. 
	
	
	
	The calculation of routing cost is a critical aspect of the detailed routing process. The routing cost for an edge \( e \in E \) is influenced by multiple factors, including traditional routing costs, stitch costs, and color conflicts. The overall routing cost \( \text{Cost}(e) \) is defined as follows:
	
	\begin{equation}
		\text{Cost}(e) = \alpha \cdot \text{Cost}_{\text{trad}}(e) + \beta \cdot \text{Cost}_{\text{stitch}}(e) + \gamma \cdot \text{Cost}_{\text{color}}(e)
	\end{equation}
	
	where:
	\begin{itemize}
		\item \( \text{Cost}_{\text{trad}}(e) \) represents the traditional routing cost associated with edge \( e \). The detailed computation is explained in ISPD contest \cite{ispd18, ispd19}.
		
		\item \( \text{Cost}_{\text{stitch}}(e) \) accounts for the cost associated with stitches, necessary when different segments of a wire are patterned using different masks and must be connected seamlessly.
		
		\item \( \text{Cost}_{\text{color}}(e) \) reflects the cost associated with color conflicts, which penalizes cases where adjacent edges share the same color state.
	\end{itemize}
	
	In this equation, \( \alpha \), \( \beta \), and \( \gamma \) are weighting factors that balance the traditional routing costs, stitch costs, and color conflict costs, respectively. 
	
	This comprehensive cost function is designed to guide the routing process in an TPL environment, optimize both the physical and manufacturability aspects of circuit design, thereby improving overall performance and reliability.
	
	\begin{algorithm}[htbp]
		\setstretch{0.85}
		\caption{Multi-pin Net Routing}
		\KwIn{The pin list of the net $pins$,
			the routing grid $G$}
		\KwOut{Path result $resPaths$}
		$startPin$ = $pins[0]$\;
		$solQueue$ = \{$\varnothing$\}\;
		$resPaths$ = \{$\varnothing$\}\;
		
		\ForEach{\textnormal{$vertice$ in $G$.get\_covered\_vertices($startPin$)}}{
			$vertice.cost$ = $0$\;
			$vertice.colorState$ = \texttt{COLORSTATE(111)}\;
			$solQueue$.push($vertice$)\;
		}
		\While{\textnormal{not all $pins$ routed}}{
			$dstVertice$ = color\_state\_searching($solQueue$, $G$)\;
			$path$ = backtrace($solQueue$, $dstVertice$)\;
			$resPaths$.push($path$)\;
		}
		\Return{$resPaths$}\;
	\end{algorithm}
	\vspace{-15pt}
	
	\subsection{Multi-pin Net Routing}

	At this stage, we search for the minimal cost path on the layout, focusing on connecting multiple pins within the net. This method incorporates color conflict costs during the path search, unlike approaches that handle color decomposition after path determination. By considering colors during pathfinding, the solution space is expanded, leading to more optimal results. Algorithm 1 provides the detailed steps.
	
	The process begins with the initial steps (lines 1–3), where the starting pin is determined, typically the first pin in the net's pin-list. An empty priority queue is initialized, using the vertice cost as the priority criterion, alongside a path list to store the routing results. Lines 4-8 involve traversing all vertices of the starting pin, setting their cost to zero, and initializing their color state with 111. These vertices are then added to the solution queue.
	
	Line 9 initiates the search for multi-pin nets by verifying whether any pins remain unrouted. If all pins are connected, the loop terminates. Line 10 performs color state searching, starting from the current vertices in the priority queue (solQueue) and continuing until an unconnected pin is found. Line 11 executes backtracing to finalize the path with its corresponding color state, as discussed in subsequent sections. Finally, Line 12 appends the completed path to the result list.
	
	\begin{algorithm}[htbp]
		\setstretch{0.85}
		\caption{Color State Searching}
		\KwIn{The priority queue of vertices $solQueue$, the routing grid $G$}
		\KwOut{The vertice of reached pin $dstVertice$}
		
		$dstVertice$ = \texttt{NULL}\;
		\While{\textnormal{$solQueue \neq \varnothing$ }}{
			$vertice$ = $solQueue$.front()\;
			\If{\textnormal{$vertice$ is covered with an unreached pin}}{
				$dstVertice$ = $vertice$\;
				break\;
			}
			\For{\textnormal{$dir\in$ directions \texttt{\{F,B,R,L,U,D\}}}}{
				$cost$ = \texttt{[[R,0],[G,0],[B,0]]}\;
				\For{\textnormal{$color\in$ colors \texttt{\{R,G,B\}}}}{
					$cost[color][0]$ = $color$\;
					$cost[color][1]$ = 
					$G$.get\_color\_cost($vertice$, $dir$, $color$)\;
					\If{\textnormal{$dir\notin$ \texttt{\{U,D\}} \&\& $color \notin vertice.colorState$}}{
						$cost[color][1]$ += $stitchCost$\;
					}
				}
				$minCost$, $colorState$ = get\_lowest($cost$)\;
				$newCost$ = $vertice.cost$ + $minCost$\;
				$solQueue$.update($vertice$, $dir$, $newCost$, $colorState$)\;
			}
		}
		\Return{$dstVertice$}
	\end{algorithm}
	\vspace{-15pt}
	
	\subsection{Color State Searching}
	
	The color state searching process is a key step in routing that dynamically evaluates and updates the routing grid to minimize overall cost while considering color conflicts and stitches. This process involves iterative calculations to ensure that paths are optimized for manufacturability within the constraints of triple patterning lithography. The algorithm takes as input the priority queue and the routing grid, iteratively searching for connections between pins. Algorithm 2 provides a detailed description of this process.
	
	Lines 1–3 initialize the necessary data structures and define the conditions for terminating the search. Lines 4–7 check whether the current vertice is covered by an unreached pin, indicating that a new pin has been located, allowing the search for that pin to end. 
	
	Line 8 initiates the search for the next vertice in all possible directions. Line 9 defines a 2-D cost array for the current direction, structured as a 3×2 matrix. The first dimension corresponds to the three possible masks, while the second dimension’s first variable stores mask information, and the second variable represents the associated color cost. Lines 10–16 handle the computation and aggregation of color costs: Lines 11–12 calculate the current cost for each mask, while Lines 13–15 evaluate whether a stitch is required and add the associated cost if present. Lines 17–19 then sort the computed color costs to identify the lowest-cost option. This minimum cost is added to vertice. cost and the updated vertice is reinserted into solQueue to continue the search.
	
	\begin{algorithm}[htbp]
		\setstretch{0.85}
		\caption{Backtrace}
		\KwIn{The priority queue of vertices $solQueue$, the vertice of reached pin $dstVertice$}
		\KwOut{Path result}
		$vertice$ = $dstVertice$\;
		\While{\textnormal{$vertice$ $\neq$ \texttt{NULL} and $vertice.cost$ $\neq$ $0$}}{
			\If {\textnormal{$vertice.verSetPtr$ = \texttt{NULL}}}{
				$vertice.verSetPtr$ = make\_verSet($vertice.colorState$)\;
				$vertice.verSetPtr.segSetPtr$ = make\_segSet($vertice.colorState$)\;
			}
			\If{\textnormal{$vertice.prev$ has conmmon color with $vertice$}}{
				\If{\textnormal{$vertice.prev.verSetPtr$ = \texttt{NULL}}}{
					$vertice.prev.verSetPtr$ = $vertice.verSetPtr$\;
				}
				\Else{
					$tState$ = get\_same\_color($vertice.prev$, $vertice$)\;
					$vertice.verSetPtr.segSetPtr.$ change\_state($tState$)\;
					$vertice.prev.verSetPtr.segSetPtr$ = $vertice.verSetPtr.segSetPtr$\;
				}
			}
			$vertice.cost$ = 0\;
			$solQueue$.update($vertice$)\;
			$vertice$ = $vertice.prev$\;
		}
		return get\_path($dstVertice$)\;
	\end{algorithm}
	\vspace{-15pt}
	
	\subsection{Backtrace}

	In this section, we introduce the backtrace algorithm, which is designed to optimize the tracing and cost evaluation of vertices along routing paths. This algorithm takes as input the priority queue of the solution and the destination vertice and outputs the finalized routing path. The core concept of the algorithm is to trace backward from the destination vertice to the starting vertice, progressively updating both the cost and color state of each vertice to ensure compliance with design rules and optimal path quality. Before describing the algorithm, it is helpful to introduce two key definitions:
	
	\hspace*{\fill}
	
	\textit{\textbf{Definition 2:}} The vertice color-set (\textit{\textbf{verSet}}) is a collection of vertices. These vertices are adjacent on the layout, are searched consecutively, and share the same color state.
	
	\textit{\textbf{Definition 3:}} The segment color-set (\textit{\textbf{segSet}}) is a collection of verSets. All versets in this set will have the same color state. Two connected vertices belong to different segSets only if stitch is introduced between them. Otherwise, adjacent vertices must belong to same segSet.
	
	\hspace*{\fill}
	
	Algorithm 3 begins by setting the current vertice to the destination vertice of the reached pin (line 1). The loop in line 2 continues until the current vertice is either null or its cost is zero. In each iteration (lines 3-6), if the verSetPtr, which points to the verSet of the vertice, of current vertice  is null, a new verSet and segSet are created based on the color state of the vertice. In lines 7-16, if the predecessor vertice shares the same color state with the current vertice, the algorithm checks if the verSetPtr of predecessor is null. If it is, the verSetPtr of predecessor is set to the verSet of current vertice. Otherwise, the algorithm computes the shared color state of both vertices and modifies the current segSet of vertice to reflect this. The segSetPtr of predecessor is then set to the current segSet of vertice. Throughout each iteration, the update function is called to update the solution of current vertice. Than, the current vertice is updated to its predecessor, and the loop continues. When the loop finishes, the final color of mask is determined by the dstVertice, which is the same as the backtracing process in Dijkstra's algorithm.
	
	\begin{table*}[!t]
		\caption{experiment results compared with \cite{dac2012} on ISPD 2018 contest}
		\centering
		\begin{threeparttable}
			\scalebox{0.89}{
				
				\begin{tabular}
					{{|l|l|l|l|l|l|l|l|l|l|l|l|l|}}
					\hline
					\multicolumn{1}{|c}{\multirow{2}{*}{case}} & \multicolumn{3}{|c}{conflict} & \multicolumn{3}{|c}{stitch} & \multicolumn{3}{|c}{cost} & \multicolumn{3}{|c|}{runtime (s)}\\
					\cline{2-13}
					& \cite{dac2012} & ours & imp. & \cite{dac2012} & ours & imp. & \cite{dac2012} & ours & imp. & \cite{dac2012} & ours & speedup \\
					\hline
					test1  & 0		& \textbf{0}		& $zero$\tnote{a}			& 3			& \textbf{0}	& 100.0\%		& 2.9545E+05	& \textbf{2.9203E+05}	& 1.15\%	& 59.93		& \textbf{14.98}   & 4.00$\times$ \\
					test2  & 0		& \textbf{0}		& $zero$			& 20		& \textbf{4}	& 80.00\%		& 4.7127E+06	& \textbf{4.6736E+06}	& 0.83\%	& 605.34		& \textbf{156.76}   & 3.86$\times$\\
					test3  & 0		& \textbf{0}		& $zero$			& 86		& \textbf{16}	& 81.40\%		& 5.3864E+06	& \textbf{5.2565E+06}	& 2.41\%	& 1932.20		& \textbf{518.25}  & 3.73$\times$\\ 
					test4\tnote{b}  & -  	& \textbf{4}		& -			& -			& \textbf{615}	& -      		& -          	& \textbf{1.5055E+07}	& -         & \(>\)24h        		& \textbf{2796.72} & -       \\  
					test5  & 2		& \textbf{0}		& 100.0\%	& 99		& \textbf{23}	& 76.77\%		& 1.6376E+07	& \textbf{1.6361E+07}	& 0.10\%	& 14188.33		& \textbf{1110.10} & 12.78$\times$\\  
					test6  & 17		& \textbf{1}		& 94.12\%	& 185		& \textbf{30}	& 83.78\%		& 2.1340E+07	& \textbf{2.1314E+07}	& 0.12\%	& 4097.95		& \textbf{886.12}  & 4.62$\times$\\ 
					test7  & 21		& \textbf{3}		& 85.71\%	& 354		& \textbf{62}	& 82.49\%		& 3.8536E+07	& \textbf{3.8489E+07}	& 0.12\%	& 14944.13		& \textbf{2272.81}  & 6.58$\times$\\ 
					test8  & 42		& \textbf{0}		& 100.0\%	& 368		& \textbf{63}	& 82.88\%		& 3.8715E+07	& \textbf{3.8674E+07}	& 0.11\%	& 12584.58		& \textbf{2143.91} & 5.87$\times$\\
					test9  & 20		& \textbf{3}		& 85.00\%	& 335		& \textbf{53}	& 84.18\%		& 3.3054E+07	& \textbf{3.3008E+07}	& 0.14\%	& 5385.06		& \textbf{1335.92} & 4.03$\times$\\  
					test10 & 352	& \textbf{274}		& 22.16\%	& 2151		& \textbf{1710}	& 20.50\%		& \textbf{4.3454E+07}	& 4.3643E+07	& -0.44\%	& 20931.53		& \textbf{6498.20} & 3.22$\times$\\  
					\hline
					avg.		   & 50.44	& \textbf{31.22}	& 81.17\%  & 400.11	& \textbf{217.89}	& 76.89\%  & 2.2430E+07	& \textbf{2.2412E+07}	& 0.51\%	& 8303.23		& \textbf{1659.68}  & 5.41$\times$\\ 
					\hline
				\end{tabular}
			}
			\begin{tablenotes}
				\footnotesize
				\item[a]{Color conflicts are all 0, so no comparison is made.}
				\item[b]{The running time of the program exceeds 24h, and no data can be obtained.}
			\end{tablenotes}
		\end{threeparttable}
	\end{table*}
	
	\begin{table}[!t]
		\caption{experiment results compared with OpenMPL\cite{openmpl} on ISPD 2019 contest}
		\centering
		\begin{threeparttable}
			\scalebox{0.87}{
				\begin{tabular}
					{{|l|l|l|l|l|l|l|}}
					\hline
					\multicolumn{1}{|c}{\multirow{2}{*}{case}} & \multicolumn{3}{|c}{conflict} & \multicolumn{3}{|c|}{stitch} \\
					\cline{2-7}
					& \cite{openmpl} & ours & imp. & \cite{openmpl} & ours & imp. \\
					\hline
					test1   & 358	& \textbf{0}	   & 100.00\%	& 279	    & \textbf{7} 	  & 97.49\%  \\
					test2   &-\tnote{a}    	& \textbf{15}	   & -			& -   	    & \textbf{388} 	  & - \\
					test3   &-    	& \textbf{2}	   & -			& -   	    & \textbf{13} 	  & - \\
					test4   &-    	& \textbf{109}	   & -			& -   	    & \textbf{1369}   & - \\
					test5   &595	& \textbf{2}	   & 99.66\%	& 303	    & \textbf{6} 	  & 98.02\% \\
					test6   &297	& \textbf{11}	   & 96.30\%	& \textbf{558}	    & 883 	  & 58.24\% \\
					test7   &-    	& \textbf{83}	   & -			& -   	    & \textbf{399} 	  & - \\
					test8   &5782	& \textbf{1}	   & 99.98\%	& 4561	    & \textbf{150} 	  & 96.71\% \\
					test9   &8966	& \textbf{130}	   & 98.55\%	& 6852	    & \textbf{234} 	  & 96.58\% \\
					test10  &10186	& \textbf{260}	   & 97.45\%	& 9433	    & \textbf{498} 	  & 94.72\% \\
					\hline
					avg. &4364	& \textbf{61.30}   & 98.66\%	& 3664.33	& \textbf{394.70} & 70.88\% \\
					\hline
				\end{tabular}
			}
			\begin{tablenotes}
				\footnotesize
				\item[a] {OpenMPL\cite{openmpl} does not provide the relevant data, the same below.}
			\end{tablenotes}
		\end{threeparttable}
	\end{table}
	
	\section{Experiment Results}
	
	We have implemented a triple patterning lithography detailed router in C++ and tested all benchmarks on a Gold 2.30 GHz Intel Xeon Linux Server with 16 cores and 512 GB memory. Mr.MPL is benchmarked against one of the most advanced detailed routing methods for triple patterning lithography \cite{dac2012}, as well as the most advanced techniques in multiple patterning lithography layout decomposition \cite{openmpl}.
	
	Mr.MPL is designed to be integrated into widely used detailed routers. To ensure a thorough comparison, we incorporated our algorithm into Dr.CU(2.0) \cite{drcu2}, one of the most advanced detailed routing systems available. Since the triple patterning lithography routing algorithm described in the \cite{dac2012} is not publicly available, we replicated it within the Dr.CU(2.0) framework to provide a fair comparison. For our experiments, we used datasets from both the ISPD 2018\cite{ispd18} and ISPD 2019 contests\cite{ispd19}, applying our multi-mask routing algorithm.
	
	\subsection{Compared with the TPL-aware Routing Method}
	
	The comparison results are presented in Table 2. The \textit{conflict} column sums the same-color and different-color conflicts. Our approach significantly reduces color conflicts, with an average decrease of 81.17\%. The \textit{stitch} column indicates the number of stitches generated during multi-mask routing, where our method achieves an average reduction of 76.89\%. The \textit{cost} column reflects the detailed routing cost score as calculated in the ISPD 2018 contest, detailed explanations can be found in the contest documentation. Our method also shows an improvement in this metric, with a 0.51\% reduction in cost.
	The final column compares the runtime of the algorithms. As anticipated from our earlier complexity analysis, our approach demonstrates a significant advantage in computational efficiency, achieving a speedup of up to 5.4$\times$. Notably, in the fourth example, \textit{ispd18test4}, the DAC 2012 method failed to produce valid results within 24h, whereas our algorithm successfully completed the task, underscoring the superiority of our approach.
	
	\subsection{Compared with the Layout Decomposition Methods}

	
	We conducted a second experiment to compare our algorithm with the most common multiple patterning lithography layout decomposition algorithm, OpenMPL. Using the dataset from the ISPD 2019 contest, we ensured a fair comparison by utilizing the data presented in the OpenMPL paper \cite{openmpl}. Additionally, the input routed layout is generated by  Dr.CU(2.0) \cite{drcu2}. The results are summarized in Table 3.
	
	The \textit{conflic} column represents the number of color conflicts, while the \textit{stitch} column indicates the number of stitches. Our method demonstrates a significant reduction in both metrics, with color conflicts and stitches decreased by 98.66\% and 70.88\%, respectively. These results strongly highlight the advantages of considering TPL during the detailed routing stage.
	
	\section{Conclusions}
	
	This paper investigates a triple patterning lithography aware detailed routing algorithm Mr.TPL. We propose a unified routing grid construction model, search, and coloring algorithm that effectively addresses TPL challenges in multi-pin nets. Our approach was compared against the state-of-the-art TPL-aware detailed routing and TPL decomposition algorithms. The experimental results demonstrate significant improvements in color conflict reduction, stitch count, cost score, and runtime, which proves the great effectiveness of our method.

	\clearpage
	\bibliographystyle{unsrt}
	\bibliography{mpl}
	
	
\end{document}